\documentclass[10pt,letterpaper]{article}
\usepackage[top=0.85in,left=1.75in,footskip=0.75in,marginparwidth=2in]{geometry}

\usepackage[utf8]{inputenc}

\usepackage{cite}

\usepackage{nameref,hyperref}

\usepackage{amsmath}
\usepackage{microtype}
\DisableLigatures[f]{encoding = *, family = * }

\raggedright
\usepackage{changepage}

\usepackage[aboveskip=1pt,labelfont=bf,labelsep=period,singlelinecheck=off]{caption}

\makeatletter
\renewcommand{\@biblabel}[1]{\quad#1.}
\makeatother

\usepackage{lastpage,fancyhdr,graphicx}
\usepackage{epstopdf}
\fancyheadoffset[L]{2.25in}
\fancyfootoffset[L]{2.25in}

\usepackage{color}

\definecolor{Gray}{gray}{.25}

\usepackage{graphicx}

\usepackage{sidecap}

\usepackage{wrapfig}
\usepackage[pscoord]{eso-pic}
\usepackage[fulladjust]{marginnote}
\reversemarginpar

\begin{document}
\vspace*{0.35in}

\begin{flushleft}
{\Large
\textbf\newline{ A study on Darboux polynomials and their significance in determining other integrability quantifiers: A case study in third-order nonlinear ordinary differential equations.}
}
\newline
\\
R. Mohanasubha 1\textsuperscript{1} and
M. Senthilvelan\textsuperscript{2}
\\
\bigskip
\textsuperscript{1} Centre for Nonlinear Systems, Chennai Institute of Technology, Chennai - 600069, Tamil Nadu, India
\\
\textsuperscript{2} Department of Nonlinear Dynamics, School of Physics, Bharathidasan University, Tiruchirappalli - 620024, Tamil Nadu, India
\\

\end{flushleft}
\begin{abstract}
 In this paper, we present a method of deriving extended Prelle-Singer method's quantifiers from Darboux Polynomials for third-order nonlinear ordinary differential equations. By knowing the Darboux polynomials and its cofactors, we extract the extended Prelle-Singer method's quantities without evaluating the Prelle-Singer method's determining equations. We consider three different cases of known Darboux polynomials. In the first case, we prove the integrability of the given third-order nonlinear equation by utilizing the Prelle-Singer method's quantifiers from the two known Darboux polynomials. If we know only one Darboux polynomial, then the integrability of the given equation will be dealt as case $2$. Likewise, case $3$ discuss the integrability of the given system where we have two Darboux polynomials and one set of Prelle-Singer method quantity. The established interconnection not only helps in deriving the integrable quantifiers without solving the underlying determining equations. It also provides a way to prove the complete integrability and helps us in deriving the general solution of the given equation. We demonstrate the utility of this procedure with three different examples.
\end{abstract}
\section{Introduction}
 Identifying the integrable systems is of great interest due to its importance and significance in the
field of science and engineering. Integrability is the unique property of these integrable systems. Though
there exist a lot of methods in the literature to deal with the
integrability of the given system, there is no unique method to study all the integrable systems. Each integrable system has to be studied and analyzed separately. These integrable systems are governed by either nonlinear partial differential equations (PDEs) or nonlinear ordinary differential equations (ODEs). To analyse the integrability of the systems governed by nonlinear
PDEs, several studies had been already reported in the
literature \cite{new_ref1, new_ref2,new_ref3,new_ref4, new_ref5, new_ref6}. For more details about the integrability of nonlinear PDEs, one can refer the references \cite{new_ref7, new_ref8, new_ref9, new_ref10, new_ref11} and therein. Likewise, methods have been developed to integrate and find the solutions of nonlinear ODEs \cite{olv,book1}. Some of the well-known mathematical methods that are being used in the recent literature to derive the solutions of nonlinear ODEs are (i) Lie symmetry analysis \cite{bluman,ibra,ste,puc}, (ii) extended Prelle-Singer procedure \cite{ps_original,annaa2,annaa3,gordoa,duar,orhan}, (iii) Darboux method \cite{darb,lim_1,lim_2,lib_3,csf2}, (iv) Jacobi last multiplier method \cite{nuc,nuc1,nuc2}, (v) $\lambda$-symmetry analysis \cite{mur_new,muri1,muri2,muri3,muri4,bhu_ml}, (vi) homogeneous balance method \cite{hpm}, its extended versions and so on \cite{o1,o2,suba4,suba5,suba6}.  Each method has its own benefits and limitations. As far as the limitations of these methods are concerned, one has to go for an ansatz in order to determine the Darboux Polynomials (DPs) in the Darboux method. Suppose the system admits a trigonometric form of integral it then becomes very difficult to obtain through the DP method. As far as the extended Prelle-Singer (PS) procedure is concerned, a proper ansatz should be made in order to determine the underlying integrable quantifiers, namely the null forms $(P, Q)$ and integrating factors $(R)$. It has been shown that certain equations are to be integrable even though they do not possess Lie point symmetries. In such circumstances, to construct the integrals, it is informative to search for more generalized symmetries. Determining those generalized symmetries is often a cumbersome task.

 Recently, efforts have been made to interlink the analytical methods \cite{muri3,ssuba1,suba2,suba3}. By interconnecting the methods, we can derive one integrability quantifier from the other. For example, suppose we know PS quantities, that is null forms and integrating factors, then the established interconnections will help in deriving the other integrability quantifiers, namely DPs, Lie symmetries, Jacobi last multipliers and  integrating factors  without recourse to the respective method. 

\par In one of our earlier works, by considering second order ODEs, we have shown a method to derive  extended PS quantities from DPs \cite{ssuba1}. We have also analysed the connection between the various analytical methods for third-order ODEs and $n^{th}$-order
ODEs in Refs. \cite{suba2} and \cite{suba3}, respectively. From the analysis, we have observed that some of the interconnections remains same for any order ODEs except their order. However, the rare connections are more complex than the corresponding lower order connections. In this paper, we create a interlink between DPs and the extended PS quantities for third order ODEs. We begin our work with DP and its cofactor at hand. Using this cofactor, we determine another function which in turn helps to identify second null form from the first null form. Finding this function plays a major role in this procedure. Determination of this function enables us to derive the other integrability quantifiers algebraically. We explain the interconnection in three categories. In the first category, we consider a situation in which we know two sets of DPs and their cofactors. Here, we aim to determine three null forms, three integrating factors and three integrals for the considered system from the known two DPs and their cofactors. Interestingly, we also explore the third DP and its cofactor through the established interconnections without solving the DP determining equation. In the second category, we consider a situation in which we know only one DP and its cofactor for the given ODE. The method of deriving the other unknown quantifiers, namely two more DPs and their cofactors and three sets of PS quantities and their associated integrals from the known quantifier will be discussed in the second case. In the third category, we consider a case in which we know two DPs and their cofactors and also one set of PS  quantities. In this case, we show that the other quantifiers and the  complete integrability of the given ODE can be established by mere algebraic calculations. The main advantage of the proposed procedure is that  instead of solving the cumbersome determining equations in the concerned method one can derive a number of  integrability quantifiers from the known  quantifier. 

\par This framework of the article is as follows: In Section 2, we recall the two analytical methods for third-order nonlinear ODEs, namely Darboux method and the extended PS procedure. In Section 3, we  connect DPs and their cofactors with extended  PS quantities. In Sec. 4, we demonstrate the method of deriving unknown quantifiers from known quantifiers for three different situations. In Sec. 5, we illustrate the procedure with suitable examples. Finally, we conclude our article in Section 6.

\section{Analytical Methods for third-order ODEs\label{s2}}
In this section, in order to be self-contained, we briefly recall the analytical methods, namely Darboux method and the extended PS procedure, which we intend to interconnect.
 
 \subsection{Darboux Method \label{2.1}}
 
Let us consider a third-order nonlinear ODE of the form
\begin{align}
\dddot{x}=\phi(t,x,\dot{x},\ddot{x}),
\label{eqn1}
\end{align}
where $\phi$ is a function of $t,~x,~\dot{x}$ and $\ddot{x}$ and overdot denotes differentiation with respect to $t$. Let us suppose that Eq. (\ref{eqn1}) admits an integral of the form $\qquad I= l_1(t, x, \dot x, \ddot x)/l_2(t,x,\dot x, \ddot x)$ where $l_1$ and $l_2$ are functions of their arguments. Upon differentiating this integral w.r.t $t$, we find  
\begin{align}
\dfrac{dI}{dt} =\dfrac{d}{dt}\dfrac{l_1}{l_2} = 0&\Rightarrow \dot{l_1}=h(t,x,\dot x,\ddot x)l_1\nonumber \\&\Rightarrow D[l_1] = h(t,x,\dot x, \ddot x)l_1.\label{didt}
\end{align}
Equation (\ref{didt}) is the determining equation for the DP \cite{darb,lim_1,lib_3}. Here $D=\frac{\partial}{\partial{t}}+
\dot{x}\frac{\partial}{\partial{x}}+
\ddot{x}\frac{\partial}{\partial{\dot{x}}}+\phi\frac{\partial}
{\partial{\ddot{x}}}$ is the total differential operator and $h(t,x,\dot{x},\ddot x)$ is the DP cofactor.  Solving (\ref{didt}), we can determine DPs ($l$) and their cofactors $(h)$ of Eq. (\ref{eqn1}). The ratio of two DPs whose cofactors are same provides an integral for Eq. (\ref{eqn1}) \cite{lim_1}. Eventhough it is sufficient to know particular solutions of (\ref{didt})  in order to explore those solutions one often assumes an ansatz for the functions $l$ and $h$.

\subsection{Extended PS procedure\label{2.2}}
\label{sec1}
In this procedure, we rewrite Eq. (\ref{eqn1}) in the form  $d\ddot x - \phi dt = 0$. Now adding the terms $Q(t,x,\dot x,\ddot x) \ddot x dt - Q(t,x,$ $\dot x,\ddot x)d\dot x$ and $P(t,x,\dot x,\ddot x) \dot x dt - P(t,x,\dot x,\ddot x)dx$ in this equation, we get
\begin{equation}
    (\phi +P\dot x +Q\ddot x) dt-P dx-Qd\dot x-d\ddot x= 0.\label{e2}
\end{equation}
Note that the terms which we appended above reshapes the third order ODE (\ref{eqn1}) into a more general $1-$form. The functions $P$ and $Q$ are called null forms. Upon multiplying Eq. (\ref{e2}) by an integrating factor ($R(t,x,\dot{x},\ddot{x})$) it can be rewritten as a perfect differential function, that is 
\begin{equation}
    R(\phi+P\dot x+Q\ddot x)dt-R Pdx-RQd\dot x-Rd\ddot x = dI = 0, \label{e3}
\end{equation}
where $I$ is an integral of Eq. (\ref{eqn1}).
\par Let us recall that the total differential $I$ of (\ref{eqn1}) can also be written as $dI = \dfrac{\partial I}{\partial t} dt+\dfrac{\partial I}{\partial x }dx+\dfrac{\partial I}{\partial\dot x }d\dot x+\dfrac{\partial I}{\partial\ddot x }d\ddot x=0.$ Comparing this total differential with the one obtained in $(\ref{e3})$, we can identify the following relations, namely
\begin{align} 
I_{t}=R(\phi+\dot{x}P+Q\ddot{x}),~I_{x}=-RP,~I_{\dot{x}}=-RQ,~I_{\ddot{x}}=-R.  
\label{met8}
\end{align} 
Upon integrating the first order ODEs (\ref{met8}), we find
\begin{align}
I(t,x,\dot{x},\ddot{x})=p_1-p_2-\int\big(RQ+\frac{d}{d \dot{x}}[p_1-p_2]\big)d\dot{x}\nonumber\\-\int \big(R+\frac{d}{d\ddot{x}}[p_1-p_2-\int [RQ+\frac{d}{d\dot{x}}[p_1-p_2]]d\dot{x}]\big)d\ddot{x}, \label{met13}
\end{align}
where
\begin{eqnarray}
 p_1 =\int R(\phi+P\dot{x}+Q\ddot{x})dt,~~
 p_2 = \int \big(RP+\frac{d}{dx}\int p_1\big)dx.\nonumber
\end{eqnarray}
\par Substituting the  expressions $P$, $Q$ and $R$ in (\ref{met13}) and integrating the resultant expressions we can obtain the  integrals of (\ref{eqn1}).

The null functions ($P$ and $Q$)  and the integrating factor  ($R$) can be determined from the integrability conditions, $I_{tx}= I_{xt}$, $I_{t\dot x}=I_{\dot x t}$, $I_{t\ddot x}=I_{\ddot x t}$, $I_{x\dot x}=I_{\dot x x}$, $I_{x\ddot x}=I_{\ddot x x}$ and $I_{\dot x \ddot x}= I_{\ddot x\dot x},$ where the first order partial derivatives of the integral $I$ are given in (\ref{met8}). These six integrability conditions provide  six equations among which three of them turns out to be the determining equations for the unknowns ($P$, $Q$ and $R$) and the other three turn out to be a  set of  constraints that have to be satisfied by these functions. The determining equations and the constraints read
\begin{eqnarray}  
D[P]=&-\phi_x+P\phi_{\ddot{x}}+QP,\label{met9}\\
D[Q]=&-\phi_{\dot{x}}+Q\phi_{\ddot{x}}-P+Q^2,\label{met91}\\
D[R]=&-R(Q+\phi_{\ddot{x}}),\label{met10}\\
R_x=&R_{\ddot{x}}P+RP_{\ddot{x}},\label{met11}\\
R_{\dot{x}}P=&-P_{\dot{x}}R+R_x Q+RQ_x,\label{met111}\\
R_{\dot{x}}=&R_{\ddot{x}}Q+RQ_{\ddot{x}}.\label{met112}
\end{eqnarray}
  The method of solving these equations has already been discussed in Ref. \cite{annaa3}.
 \par Let us suppose that  three independent solutions of Eqs. (\ref{met9})-(\ref{met112}) be $P_i$, $Q_i$, $R_i$, $i=1,2,3$. For each set we can construct an integral through (\ref{met13}). Let us designate these integrals be 
\begin{eqnarray}
I_1&=&N_1(t,x,\dot{x},\ddot{x}),\label{ei1}\\
I_2&=&N_2(t,x,\dot{x},\ddot{x}),\label{11}\\
I_3&=&N_3(t,x,\dot{x},\ddot{x})\label{ei3}.
\label{eq:}
\end{eqnarray}
Thus the complete integrability of the given equation (\ref{eqn1}) can be established through the extended PS procedure by determining three sets of null forms ($P$ and $Q$) and the integrating factors ($R$).


\section{Extracting one integrability quantifier from another known quantifier\label{3}}
Now, we connect DPs with the PS quantities. By succeeding this task, we can construct the integrability quantifiers of the second method from the integrability quantifiers of the first method itself. 
Rearranging the expression (\ref{ei1}) for $\ddot{x}$ and substituting it into (\ref{11}), we find 
\begin{eqnarray}
I_2=\tilde{N_2}(t,x,\dot{x},I_1).
\label{eq2}
\end{eqnarray}
Differentiating Eq.(\ref{eq2}) w.r.t $x$, $\dot{x}$ and $\ddot{x}$ separately, we obtain
\begin{eqnarray}
I_{2x}&=&\tilde{N}_{2x}+\tilde{N}_{2I_1}I_{1x}=-p_2P_2,\label{eq11}\\
I_{2\dot{x}}&=&\tilde{N}_{2\dot{x}}+\tilde{N}_{2I_1}I_{1\dot{x}}=-p_2Q_2,\label{eq22}\\
I_{2\ddot{x}}&=&\tilde{N}_{2I_1}I_{1\ddot{x}}= p_2\label{eq33},
\label{eq:}
\end{eqnarray}
where we have considered the expressions given in (\ref{met8}) in order  to obtain the right hand side of the above Eqs. (\ref{eq11})-(\ref{eq33}).
Replacing the terms $I_{1\ddot{x}}$ and $I_{2\ddot{x}}$ by $p_1$ and $p_2$ respectively in Eq. (\ref{eq33}), we find
\begin{eqnarray}
p_2=l_1 p_1, \qquad l_1=\tilde{N}_{2I_1}.
\label{r2expression}
\end{eqnarray}
 \par  Substituting the relations, $I_{1x}=-p_1P_1$ and $I_{2\ddot{x}}=\tilde{N}_{2I_1}I_{1\ddot{x}}=p_2$, given in (\ref{met8}) in (\ref{eq33}) and simplifying the resultant equation we arrive at the following expression for the second null form $P_2$, that is 
\begin{eqnarray}
P_2=m_1+P_1,
\label{s2expression}
\end{eqnarray}
where the unknown function $m_1(=\frac{\tilde{N}_{2x}}{\tilde{N}_{2I_1} I_{1\ddot{x}}})$ is to be determined. The expression (\ref{s2expression}) connects the first null form $P_1$ with second null form $P_2$.
\par Let us consider Eq. (\ref{eq22}) and replace the terms  $I_{1\dot{x}}$ by $-p_1Q_1$ and $p_2$ by $-\tilde N_{2I_1}I_{1\ddot{x}}$. This action yields the following expression for $Q_2$, namely
\begin{eqnarray}
Q_2=c_1+Q_1,
\label{u2expression}
\end{eqnarray}
where the function  $c_1(=\frac{\tilde{N}_{2\dot{x}}}{\tilde{N}_{2I_1} I_{1\ddot{x}}})$ is yet another function. The expression (\ref{u2expression}) connects the first null form $Q_1$ with the second null form $Q_2$. The relations (\ref{r2expression}), (\ref{s2expression}) and (\ref{u2expression}) provides a way to  determine the second integrating factor ($p_2$) and the second set of null forms ($P_2$ and $Q_2$) from the quantities ($P_1, Q_1$ and $p_1$) by determining the functions $m_1$, $c_1$ and $l_1$. The functions $m_1$, $c_1$ and $l_1$ are functions of $t, x, \dot x$ and $\ddot x$. Now we lay out a procedure to determine these three functions.

Let us differentiate Eq. (\ref{s2expression}) w.r.t $t$ and substitute it into (\ref{met9}). In this process we replace  the functions $Q_2$ and $P_2$  by the expressions found in  (\ref{s2expression}) and (\ref{u2expression}). Implementing this, we end up at
\begin{eqnarray}
D[m_1]=m_1(\phi_{\ddot{x}}+c_1+Q_1)+c_1P_1.
\label{b1_eqn}
\end{eqnarray}
A similar equation can also be derived for the unknown function $c_1$ by  differentiating Eq. (\ref{u2expression}) w.r.t $t$ and substituting the resultant expression along with (\ref{s2expression}) and (\ref{u2expression}) in the PS method determining equation (\ref{met91}). Here, we find 
\begin{eqnarray}
D[c_1]=c_1^2+c_1(\phi_{\ddot{x}}+2Q_1)-m_1.
\label{c1_eqnrr1}
\end{eqnarray}
On the other hand differentiating  Eq. (\ref{r2expression}) with respect to $t$ and substituting the determining equation (\ref{met10}) into it, we end up at
\begin{eqnarray}
D[l_1]=-c_1l_1.\label{22}
\label{darboux}
\end{eqnarray}
Equation (\ref{22}) is nothing other than the determining equation for DP (see Eq. (\ref{didt})) in which $l_1$ is the DP and $(-c_1=h)$ is the cofactor. We note that the third set of quantifiers ($P_3, Q_3$, $R_3$) can  be determined using the same expressions (\ref{r2expression}), (\ref{s2expression}) and (\ref{u2expression}) by considering $P_2$, $Q_2$ and $p_2$ as known quantities and $P_3$, $Q_3$ and $R_3$ as unknowns.

\section{Methodology to derive the integrability of the third-order ODEs}
\subsection{Case 1: Two DPs are known\label{3.2}}

To begin, let us assume that we know two DPs ($l_1$ and $l_2$)  and their associated cofactors ($c_1$ and $c_2$) of Eq. (\ref{eqn1}). These two polynomials are solutions of Eq. (\ref{22}), that is $D[l_i]=c_il_i,~ i=1,2$.   Now we determine the PS quantities, $P_i$, $Q_i$, $R_i$, $i=1,2,3,$ and their integrals from the two known DPs.  
  \par Substituting the known expression $c_1$  in Eq. (\ref{c1_eqnrr1}), we express the null form $Q_1$ in terms of the function $m_1$. Inserting this expression into (\ref{b1_eqn}), we get an equation which involves $P_1$ and $m_1$. Now plugging this expression into (\ref{met9}) and solving the resultant differential equation, we obtain an explicit form of $m_1$.
  \par Substituting the determined function $m_1$ back in (\ref{c1_eqnrr1}) we can obtain the form of $Q_1$. Inserting this null form $Q_1$ in (\ref{met91}) we can get the null form $P_1$. The integrating factor $p_1$ can be calculated from the last expression given in  (\ref{met10}). With the known expressions $P_1$ and $m_1$, Eq. (\ref{s2expression}) yields the expression for second null form $P_2$. Since $c_1$ and $Q_1$ are known, inserting them in (\ref{u2expression}) helps us in  identifying $Q_2$. The second integrating factor $p_2$ can be obtained from the expression (\ref{r2expression}). In this way, we can generate two sets of PS quantities, namely  $U_i$, $S_i$ and $R_i$, $i=1,2,$ from the known DP $l_1$ and its cofactor  $c_1$. We note that one can also reformulate the steps given above and alternatively derive the quantifiers $U_i$, $S_i$, and $R_i$, $i=1,2,$. This essentially  depends upon the difficulties one may come across while following the steps given before. However, in either of the methods one has to solve only one differential equation. The rest of the calculations involve only algebra.

\par To explore the third null form $Q_3$, we reconsider the expression (\ref{u2expression}) in the form
\begin{eqnarray}
Q_3=c_2+Q_2.
\label{u3expression}
\end{eqnarray} 
Here $c_2$ is the cofactor of second DP $(l_2)$ and $Q_2$ is the second null form. Since both are known the third null form $Q_3$ can be identified  from (\ref{u3expression}). To determine the other null form $P_3$, we consider Eq. (\ref{s2expression}) in the form
\begin{eqnarray}
P_3=m_2+P_2.
\label{s3expression}
\end{eqnarray}
Following our earlier foot steps, the function $m_2$ can be determined from Eq. (\ref{c1_eqnrr1}) by considering it in the form $D[c_2]= c_2^2+c_2(\phi_{\ddot{x}}+2Q_2)-m_2$. Plugging the known functions $c_2$ and $Q_2$ in this equation and rewriting, we can obtain the expression for $m_2$. Since the functions $m_2$ and $P_2$ are known we can fix the third null form $P_3$. The integrating factor $R_3$ can be fixed from the relation
\begin{eqnarray}
R_3=l_2p_2.
\label{r3expression}
\end{eqnarray}
Substituting each set of the functions $P_i$, $Q_i$ and $R_i$, $i=1,2,3,$ separately in (\ref{met13}) and evaluating the integrals we can obtain the first integrals of Eq. (\ref{eqn1}). 
\par Thus, knowledge of the DPs  and their cofactors help to identify the integrability quantifiers and the integrals for the given ODE. One may observe that the function $Q_3$ can be obtained from $Q_2$ only the cofactor $c_2$ associated with the second DP is known. In some circumstances, we may know only one DP and its cofactor for the given equation. In this case, we  adopt another way and determine the second DP ($l_2$) and its cofactor ($c_2$). We investigate this situation as a separate case in the following.

\subsection{Case 2: Only one  DP is known\label{3.3}}
 Let us suppose that we know only one DP ($l_1$) and its associated cofactor ($c_1$). In this case we need to determine the second DP $l_2$  and its cofactor $c_2$ in order to capture other integrability quantifiers. To succeed in this case we proceed as follows. 

\par We start the procedure from Eq. (\ref{r3expression}). Substituting Eq. (\ref{r2expression}) into it we find $R_3=l_2l_1p_1$ from which we can obtain a relation
\begin{eqnarray}
l_2l_1=\frac{R_3}{p_1}.
\label{k1k2}
\end{eqnarray}
Since $l_1$ and $l_2$ are two DPs their product ($l_1l_2$) can also be considered as a DP \cite{lib_3}. Since the L.H.S in (\ref{k1k2}) is a polynomial the R.H.S
 $\left(R_3/p_1\right)$ may also be considered as a polynomial. Considering this fact we identify the following relation, that is 
\begin{eqnarray}
l_2=\frac{f}{l_1},
\label{k2new}
\end{eqnarray}
where $f=\frac{R_3}{p_1}$. Since the function $f$ is a DP it should satisfy the equation 
\begin{eqnarray}
D[f]=gf,
\label{feqn}
\end{eqnarray}
with $g$ is the associated cofactor. 
\par Differentiating Eq. (\ref{k2new}) with respect to $t$ and replacing the terms $D[l_2]$ by $-c_2l_2$ and $D[f]$ by $gf$  and simplifying the resultant equation, we find
\begin{eqnarray}
c_2=-(c_1+g).
\label{eq:c2}
\end{eqnarray}
Substituting (\ref{eq:c2}) in the determining equation for $c_2$ (refer Eq. (\ref{c1_eqnrr1})) the latter equation becomes
\begin{align}
D[c_1]+D[g]=-(c_1^2&+g^2+2gc_1)+(c_1+g)(\phi_{\ddot{x}}+2Q_2)+m_2.
\label{eqgh}
\end{align}
The unknown functions in Eq. (\ref{eqgh}) are $g$ and $m_2$.  Considering Eq. (\ref{b1_eqn}) for $m_2$ in the form 
$D[m_2]=m_2(\phi_{\ddot{x}}+c_2+Q_2)+c_2P_2 $ and substituting Eq. (\ref{eq:c2}) into it and rearranging the resultant equation, we end up with  
\begin{eqnarray}
D[m_2]=m_2(\phi_{\ddot{x}}-c_1-g+Q_2)-(c_1+g)P_2.
\label{eqfh}
\end{eqnarray}

Inserting the  expression  $m_2$ from  (\ref{eqgh}) in the above Eq. (\ref{eqfh}), the later equation turns out to be the determining equation for $g$. Solving this equation we can determine the explicit form of $g$, from which we can identify the function $m_2$ through (\ref{eqgh}). The DP $f$ can be found  by substituting the expression $g$ back in Eq. (\ref{feqn}) and solving the resultant equation. In this way we can identify not only the second DP $l_2$ and its cofactor $g$ but also the needed functions $m_2$ and $c_2$. The third DP, its cofactor and the extended PS procedure quantities can be derived in the same way as we did in the previous case $1$.

\subsection{Case 3: Two  DPs and first set of PS method quantities are known\label{3.3}}
Suppose we know two DPs, their cofactors and the first set of PS method quantities, then the remaining quantifiers of the considered system can be determined algebraically. By substituting the cofactor $c_1$ and the null form $Q_1$ in (\ref{c1_eqnrr1}) and simplifying the resultant expression, we can obtain the function $m_1$. Substituting the DP $l_1$, its cofactor $c_1$, the function $m_1$ and the first set of PS  quantities in the expressions (\ref{r2expression}), (\ref{s2expression}) and (\ref{u2expression}) and simplifying the resultant equations, we can  identify the second set of PS  quantities $(P_2,Q_2,p_2)$. Now repeating the same procedure with the second set PS  quantities $(P_2,Q_2,p_2)$, DP $(l_2)$  and its cofactor $(c_2)$, we can derive the third set of PS method quantities. The function $m_2$ can be derived through the expression (vide Eq. (\ref{c1_eqnrr1}))
\begin{eqnarray}
D[c_2]=c_2^2+c_2(\phi_{\ddot{x}}+2Q_2)-m_2.
\label{c2_eqnrr1}
\end{eqnarray}
The associated integrals can be constructed with the help of (\ref{met13}).

\section{Illustrations for the procedure}
In this section,  we consider three examples and apply the theory developed in the previous section to these examples. In the first example, we consider the situation in which two DPs are known and in the second example we consider the situation where only one DP is known. In the third example, we consider a situation in which we have partial information on DP and partial information on PS quantifiers.  With these partial information at hand, in the third example, we demonstrate the complete integrability of the given equation can be  established in an algebraic manner through the interconnections found in this work.
\subsection{Example 1}
We consider Chazy equation of the form \cite{annaa3}
\begin{eqnarray}
\dddot{x}+4 x \ddot{x} + 3 \dot{x}^2 + 6 x^2 \dot{x} + x^4=0.\label{32}
\label{eqdddx}
\end{eqnarray}
Equation (\ref{32}) admits two DPs of the form
\begin{align}
l_1&=-\frac{t (x (t x-2)+t {\dot{x}}) \left(x^3+3 x {\dot{x}}+{\ddot{x}}\right)^2}{\left(x^2+{\dot{x}}\right) \left(-t x^3-3 t x {\dot{x}}-t {\ddot{x}}+x^2+{\dot{x}}\right)^2},\label{eqK1}\\
l_2&=\frac{t^2 \left(x^2+{\dot{x}}\right)-3 t x+3}{3 (x (t x-2)+t {\dot{x}})}. 
\label{Dark2_ex1}
\end{align}
The associated cofactors are given by
\begin{align}
c_1&=\frac{2 x \left(-t x^3-3 t x {\dot{x}}-t {\ddot{x}}+x^2+{\dot{x}}\right)}{t \left(x^2+{\dot{x}}\right) (x (t x-2)+t {\dot{x}})},\\
c_2&=-\frac{(t x-3) \left(t x^3+3 t x {\dot{x}}+t {\ddot{x}}-x^2-{\dot{x}}\right)}{(x (t x-2)+t {\dot{x}}) \left(t^2 \left(x^2+{\dot{x}}\right)-3 t x+3\right)}.
\label{eqc2}
\end{align}
One can check that the expressions given in (\ref{eqK1})-(\ref{eqc2}) satisfy Eq. (\ref{didt}). In the following, we derive the PS quantifiers and establish the integrability of (\ref{eqdddx}).
\par Substituting the cofactor $c_1$ in (\ref{b1_eqn}) the latter equation becomes
\begin{align}
D[m_1]-m_1\bigg(\frac{2 x \left(-t x^3-3 t x {\dot{x}}-t {\ddot{x}}+x^2+{\dot{x}}\right)}{t \left({\dot{x}}+x^2\right) (t {\dot{x}}+x (t x-2))}-4x+Q_1\bigg)\nonumber\\-\bigg(\frac{2 x \left(-t x^3-3 t x {\dot{x}}-t {\ddot{x}}+x^2+{\dot{x}}\right)}{t \left({\dot{x}}+x^2\right) (t {\dot{x}}+x (t x-2))}\bigg)P_1=0.
\label{37}
\end{align}
Rewriting the above Eq. (\ref{37}) for the null form $Q_1$, we find
\begin{align}
Q_1=\frac{1}{m_1}\bigg[D[m_1]-m_1\bigg(\frac{2 x \left(-t x^3-3 t x {\dot{x}}-t {\ddot{x}}+x^2+{\dot{x}}\right)}{t \left({\dot{x}}+x^2\right) (t {\dot{x}}+x (t x-2))}-4x\bigg)\nonumber \\-\bigg(\frac{2 x \left(-t x^3-3 t x {\dot{x}}-t {\ddot{x}}+x^2+{\dot{x}}\right)}{t \left({\dot{x}}+x^2\right) (t {\dot{x}}+x (t x-2))}\bigg)P_1\bigg].
\label{eq42}
\end{align}
Substituting this expression into Eq. (\ref{met91}) and rewriting it, we obtain an evolution equation  for $P_1$ in terms $m_1$. Now plugging this expression into (\ref{met9}) we can obtain a determining equation for the unknown $m_1$. Since the resultant expression is lengthy we do not reproduce the equation here. Equation (\ref{eq42}) admits a particular solution of the form
	\begin{eqnarray}
	m_1=-\frac{2 \left(x^2-{\dot{x}}\right) \left(t x^3+3 t x {\dot{x}}+t {\ddot{x}}-x^2-{\dot{x}}\right)}{t \left(x^2+{\dot{x}}\right) (x (t x-2)+t {\dot{x}})}.\label{39}
\end{eqnarray}
\par To determine the null form $Q_1$, we substitute the functions $m_1$ and $c_1$ back in Eq. (\ref{c1_eqnrr1}) which in turn yields the following expression for the null form $Q_1$, that is 
\begin{eqnarray}
Q_1=\frac{2 x^3-{\ddot{x}}}{x^2+{\dot{x}}}.
\label{u1_ex1}
\end{eqnarray}
Substituting this expression in Eq. (\ref{met91}) and simplifying it, we find 
\begin{eqnarray}
P_1=-\frac{{\ddot{x}}}{x}.
\label{s1_ex1}
\end{eqnarray}
To determine the integrating factor $p_1$, we utilize Eq. (\ref{met10}).  Upon solving this equation  we find 
\begin{eqnarray}
p_1=\frac{x^2+{\dot{x}}}{\left(x^3+3 x {\dot{x}}+{\ddot{x}}\right)^2}.
\label{r1_ex1}
\end{eqnarray}
Thus from the knowledge of one DP ($l_1$) and its associated cofactor ($c_1$)  we can obtain the PS quantifiers ($Q_1$, $P_1$, $p_1$). 
\par Now inserting the functions  $P_1$, $p_1$ and  $Q_1$ in (\ref{met13}) and evaluating the integrals we find an integral of (\ref{32}) in the from 
\begin{equation}
    I_1=\frac{x^2+{\dot{x}}}{x^3+3 x {\dot{x}}+{\ddot{x}}}-t.
\end{equation}
 \par Now we proceed to calculate the second set of null forms $(P_2, Q_2)$ and the integrating factor $(p_2)$  from the known expressions $m_1, c_1, l_1, Q_1, P_1$ and $p_1$ (vide Eqs. (\ref{s2expression}), (\ref{u2expression}) and (\ref{r2expression})). Following the procedure given in Subsec. 4 we obtain the following expression for $P_2, Q_2, p_2$, that is
\begin{align}
P_2&=\frac{t^2 x^4-2 t^2 x {\ddot{x}}+{\dot{x}} \left(3 t^2 {\dot{x}}-2\right)-4 t x^3+2 t {\ddot{x}}+2 x^2}{t (x (t x-2)+t {\dot{x}})},\label{44}\\
Q_2&=\frac{2 x (t x (t x-3)+1)-t^2 {\ddot{x}}}{t (x (t x-2)+t {\dot{x}})},\label{45}\\
p_2&=-\frac{t (x (t x-2)+t {\dot{x}})}{\left(-t x^3-3 t x {\dot{x}}-t {\ddot{x}}+x^2+{\dot{x}}\right)^2}.\label{46}
\end{align}
Inserting these expressions (\ref{44})-(\ref{46}) in (\ref{met13}) and evaluating the integrals, we identify the second integral of Eq.  (\ref{32}) in the form 
\begin{equation}
 I_2=\frac{t^2 \left(-\left(x^3+3 x{\dot{x}}+{\ddot{x}}\right)\right)+2 t \left(x^2+{\dot{x}}\right)-2 x}{-t \left(x^3+3 x {\dot{x}}+{\ddot{x}}\right)+x^2+{\dot{x}}}.
    \label{I2}
\end{equation}
 \par To obtain the third set of integrable quantifiers ($P_3,$ $Q_3, R_3$), we need to determine the  function $m_2$. To do so,  we consider Eq. (\ref{c1_eqnrr1}) in the form
\begin{eqnarray}
D[c_2]=c_2^2+c_2(\phi_{\ddot{x}}+2Q_2)-m_2.
\label{c11_eqn}
\end{eqnarray}
Inserting the known null form $Q_2$ and the DP cofactor $c_2$ in (\ref{c11_eqn}) and rearranging the resultant equation for $m_2$, we find
\begin{align}
m_2=-\frac{(t (x (t x-6)-t {\dot{x}})+6) \left(t x^3+3 t x {\dot{x}}+t {\ddot{x}}-x^2-{\dot{x}}\right)}{t (x (t x-2)+t {\dot{x}}) \left(t^2 \left(x^2+{\dot{x}}\right)-3 t x+3\right)}.
\label{eq:}
\end{align}
With the help of the functions $m_2$ and $c_2$, the third set of null forms $P_3, Q_3$ and the integrating factor $R_3$ can be captured through the expressions (\ref{s3expression}), (\ref{u3expression}) and (\ref{r3expression}). Our results show
{\small
\begin{align}
P_3&=\frac{t^3 x^4-2 x \left(t^3 {\ddot{x}}+3\right)-6 t^2 x^3+3 t \left({\dot{x}} \left(t^2 {\dot{x}}+2\right)+t {\ddot{x}}\right)+12 t x^2}{t \left(t^2 \left(x^2+{\dot{x}}\right)-3 t x+3\right)},\\
Q_3&=\frac{t^3 (-{\ddot{x}})+t x (t x (2 t x-9)+12)-3}{t \left(t^2 \left(x^2+{\dot{x}}\right)-3 t x+3\right)},\\
R_3&=-\frac{t \left(t^2 \left(x^2+{\dot{x}}\right)-3 t x+3\right)}{3 \left(-t x^3-3 t x {\dot{x}}-t {\ddot{x}}+x^2+{\dot{x}}\right)^2}.
\label{eq:}
\end{align}}
This third set  of null forms ($P_3$, $Q_3$) and the integrating factor ($R_3$) helps to construct the third integral of Eq. (\ref{32}) and the resultant form becomes 
	\begin{equation}
	  I_3=\frac{t^3 \left(-\left(x^3+3 x {\dot{x}}+{\ddot{x}}\right)\right)+3 t^2 \left(x^2+{\dot{x}}\right)+6 (1-t x)}{6 \left(-t \left(x^3+3 x {\dot{x}}+{\ddot{x}}\right)+x^2+{\dot{x}}\right)}.\label{I3}  
	\end{equation}

	The general solution of (\ref{eqdddx}) can be derived using the  integrals  $I_1$, $I_2$ and $I_3$ and it is given by 
	\begin{eqnarray}
	x(t)=\frac{\frac{t^2}{2}+I_1 t+I_1 I_2}{\frac{t^3}{6}+I_1\frac{t^2}{2}+I_1I_2t+I_1I_3}.
	\label{eq:}
	\end{eqnarray}
\begin{figure}[ht!]
\centering
\includegraphics[width=0.4\textwidth]{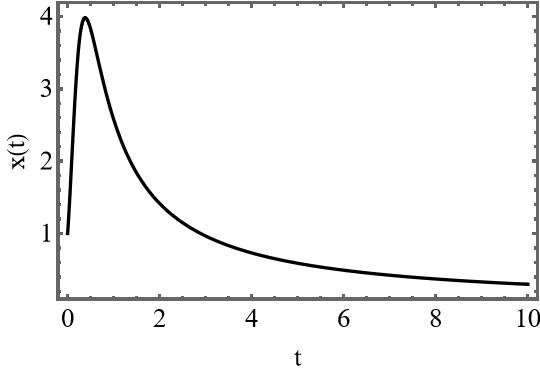}
\caption{Solution plot for Eq. (\ref{32}) with the initial conditions $I_1=0.1,\;I_2=0.1,\;I_3=0.1$.}
\label{fig:msf_sig1_sig2_yy_0.3hr}
\end{figure}
\subsection{Example 2}
In the second example, we consider a situation in which we know only one DP and its cofactor. Now the task is to determine the remaining two sets of DPs and other integrability quantifiers used in the PS procedure. To demonstrate this, we consider an equation of the form \cite{bluman}
\begin{eqnarray}
\dddot{x}=\frac{6 t {\ddot{x}}^3}{{\dot{x}}^2}+\frac{6 {\ddot{x}}^2}{{\dot{x}}}.
\label{eq59}
\end{eqnarray}
The above equation admits a DP of the form
\begin{eqnarray}
l_1={\dot{x}},
\label{eq:}
\end{eqnarray}
with the cofactor 
\begin{eqnarray}
c_1=-\frac{{\ddot{x}}}{{\dot{x}}}.
\label{eq:}
\end{eqnarray}
\par In the following, we adopt the procedure given in Subsection \ref{3.2} and obtain the null forms and its integrating factors. Now following the steps given in the previous example (vide Eqs. (\ref{37}) - (\ref{r1_ex1})) we can deduce  the first set of null form and integrating factor in the form
\begin{eqnarray}
P_1=\frac{2 {\ddot{x}}^2}{{\dot{x}}^2},\label{60}\quad
Q_1=\frac{-6 t {\ddot{x}}^2-2 {\dot{x}} {\ddot{x}}}{{\dot{x}}^2},\quad
p_1=\frac{{\dot{x}}^2}{{\ddot{x}}^2}.
\label{eq:}
\end{eqnarray}
While deriving the above functions, we also come across the following expression for $m_1$, that is 
\begin{eqnarray}
m_1=-\frac{2 {\ddot{x}}^2}{{\dot{x}}^2}.
\label{63}
\end{eqnarray}
Using (\ref{60}) and (\ref{63}), we can obtain the second set of null forms $(P_2, Q_2)$ and the integrating factor $(p_2)$ with the help of Eqs. (\ref{s2expression}), (\ref{u2expression}) and (\ref{r2expression}). The resultant outcome shows that
\begin{eqnarray}
P_2=0,\quad
Q_2=\frac{-6 t {\ddot{x}}^2-3 {\dot{x}} {\ddot{x}}}{{\dot{x}}^2},\quad
p_2=\frac{{\dot{x}}^3}{{\ddot{x}}^2}.
\label{eq:}
\end{eqnarray}
\par To obtain  the third set of null forms ($P_3$, $Q_3$) and the integrating factor ($R_3$), we need to determine the second DP ($l_2$) and its cofactor ($c_2$) from the first DP $l_1$. 
\par Substituting the known expressions $c_1$ and $Q_2$ in  Eq. (\ref{eqgh}) and simplifying it, we obtain the following determining equation for $g$, i.e.,
\begin{align}
D[g]+g^2-2g\bigg(\frac{ \ddot{x} (3 t\ddot{x}+4 \dot{x})}{\dot{x}^2}\bigg)-m_2+2 \frac{\ddot{x}^2}{\dot{x}^3} (5 \dot{x} + 6 t\ddot{x})=0.
\label{ghnj}
\end{align}
\par Rewriting the above expression for $m_2$ and substituting this expression in (\ref{eqfh}) we obtain a differential equation in terms of the  unknown function $m_2$. To proceed further, we choose a  trivial solution for $m_2$, that is
\begin{eqnarray}
m_2=0.
\label{eq:}
\end{eqnarray}
Substituting this trivial form in Eq. (\ref{ghnj}) and solving the resultant expression, we identify the following particular solution for the function $g$,
\begin{eqnarray}
g=\frac{2 \ddot{x}}{\dot{x}}.
\label{eq:}
\end{eqnarray}
Plugging this expression in (\ref{feqn}) and solving the latter equation, we find a particular solution for $f$ as
\begin{eqnarray}
f=\dot{x}^2.
\label{eq:}
\end{eqnarray}
Since the functions $f$ and $l_1$ are now determined the second DP $l_2$ can be readily identified from $l_2=\frac{f}{l_1}$, which in turn reads
\begin{eqnarray}
l_2={\dot{x}}.
\label{eq:}
\end{eqnarray}
We  note here that the second DP is same as first DP $l_1$. The associated cofactor turns out to be $c_2=-\frac{{\ddot{x}}}{{\dot{x}}}$. 
\par Equations (\ref{u3expression})-(\ref{r3expression}) yield the third set of null forms and integrating factor in the form
\begin{eqnarray}
P_3=0,\quad
Q_3=\frac{-6 t {\ddot{x}}^2-4 {\dot{x}} {\ddot{x}}}{{\dot{x}}^2},\quad
R_3=\frac{{\dot{x}}^4}{{\ddot{x}}^2}.
\label{eq:}
\end{eqnarray}

The functions 
$P_i$, $Q_i$, $R_i$, $i=1,2,3,$ help in building the necessary integrals of (\ref{eq59}) whose explicit forms are given by
\begin{align}
I_1=6 t {\dot{x}}-2 x+\frac{{\dot{x}}^2}{{\ddot{x}}},\quad
I_2=3 t {\dot{x}}^2+\frac{{\dot{x}}^3}{{\ddot{x}}},\quad
I_3=2 \left(2 t {\dot{x}}^3+\frac{{\dot{x}}^4}{{\ddot{x}}}\right).
\label{eq:}
\end{align}
Using the integrals $I_1, I_2$ and $I_3$  we can derive an implicit solution for Eq. (\ref{eq59}) in the form 
\begin{align}
I_2((2x+I_1)^2-12tI_2)^2+3t(9tI_3+I_2(I_1-2x))^2\nonumber\\
-(2x+I_1)((2x+I_1)^2-12tI_2)(9tI_3+I_2(2x+I_1))=0.
\end{align}
\subsection{Example 3}
Here, we consider a situation in which we have partial information on DPs and also have partial information on PS quantifiers. In this case,  the procedure developed in this article supports to study the complete integrability of the considered equation in an algebraic manner, as shown below.
\par Let us consider a third order nonlinear ODE, of the form \cite{euler}
\begin{eqnarray}
\dddot{x}=\frac{{\dot{x}} {\ddot{x}}}{x}+\frac{{\ddot{x}}^2}{{\dot{x}}}.
\label{eq81}
\end{eqnarray}
Suppose that we know two DPs of this equation, say
\begin{eqnarray}
l_1&=&2 {\dot{x}}-\frac{2 x {\ddot{x}}}{{\dot{x}}},\\
l_2&=&\frac{x {\dot{x}} \left(t {\dot{x}}^2-x (t {\ddot{x}}+{\dot{x}})\right)}{4 \sqrt{-x {\ddot{x}} \left(x {\ddot{x}}-2 {\dot{x}}^2\right)} \left(x {\ddot{x}}-{\dot{x}}^2\right)}.
\label{eq:}
\end{eqnarray}
and their corresponding cofactors 
\begin{eqnarray}
c_1&=&\frac{{\dot{x}} {\ddot{x}}}{{\dot{x}}^2-x {\ddot{x}}},\\
c_2&=&-\frac{x {\ddot{x}} \left(x {\ddot{x}}-2 {\dot{x}}^2\right)}{\left({\dot{x}}^2-x {\ddot{x}}\right) \left(t {\dot{x}}^2-x (t {\ddot{x}}+{\dot{x}})\right)}.
\label{eq:}
\end{eqnarray}
\par Let us also assume that we know the  null forms ($P_1, Q_1$) and the  integrating factor ($p_1$) of Eq. (\ref{eq81}), say 
\begin{eqnarray}
P_1= -\frac{{\ddot{x}}}{x}, \quad 
Q_1=-\frac{{\ddot{x}}}{{\dot{x}}},\quad
p_1=-\frac{1}{x {\dot{x}}}.
\label{eq86}
\end{eqnarray}
Now we deduce the other quantifiers in an algebraic way as follows. 
\par Since $c_1$ and $Q_1$ are known $m_1$ can be fixed from  (\ref{c1_eqnrr1}). Here, we find 
\begin{equation}
  m_1=\frac{{\ddot{x}}^2}{x {\ddot{x}}-{\dot{x}}^2}.  
\end{equation}
Substituting the expression (\ref{eq86}) in Eqs. (\ref{s2expression}), (\ref{u2expression}) and (\ref{r2expression}) the expression , we can obtain the following quantifiers, namely
\begin{eqnarray}
P_2=\frac{{\dot{x}}^2 {\ddot{x}}}{x \left(x {\ddot{x}}-{\dot{x}}^2\right)},\quad
Q_2=\frac{x {\ddot{x}}^2}{{\dot{x}}^3-x {\dot{x}} {\ddot{x}}},\quad
p_2=\frac{2 {\ddot{x}}}{{\dot{x}}^2}-\frac{2}{x}.
\label{eq:}
\end{eqnarray}
To obtain the third set of integrable quantifiers $P_3, Q_3$ and $R_3$, we have to determine the function $m_2$. It can be derived from (\ref{c11_eqn}) through a direct path. The function $m_2$ turns out to be 
\begin{eqnarray}
m_2=-\frac{x {\ddot{x}} \left(x {\ddot{x}}-2 {\dot{x}}^2\right)}{\left({\dot{x}}^2-x {\ddot{x}}\right) \left(t {\dot{x}}^2-x (t {\ddot{x}}+{\dot{x}})\right)}.
\label{eq:}
\end{eqnarray}
We can capture  the third set of integrable quantifiers with the help of Eqs. (\ref{u3expression})-(\ref{r3expression}). The resultant expressions read
\begin{eqnarray}
P_3&=&\frac{{\dot{x}} {\ddot{x}} (t {x+\dot{x}})}{x \left(x ({\dot{x}}+t {\ddot{x}})-t {\dot{x}}^2\right)},\quad
Q_3=\frac{x {\ddot{x}} (t {\ddot{x}}+2 {\dot{x}})}{t {\dot{x}}^3-x {\dot{x}} (t {\ddot{x}}+{\dot{x}})},\nonumber\\
R_3&=&\frac{t \dot{x}^2-x (t \ddot{x}+\dot{x})}{2 \dot{x} \sqrt{x \ddot{x} \left(2 \dot{x}^2-x \ddot{x}\right)}}
\label{eq:}
\end{eqnarray}
The integrals can be constructed from Eq. (\ref{met13}) by appropriately substituting the functions $P_i$, $Q_i$, $R_i$, $i=1,2,3,$ into it and the integrals are
\begin{align}
I_1&=\frac{{\ddot{x}}}{x {\dot{x}}},\;\;\;
I_2={\ddot{x}} \left(\frac{2}{x}-\frac{{\ddot{x}}}{{\dot{x}}^2}\right),\nonumber\\
I_3&=-\frac{t} {2\dot{x}}\sqrt{\frac{\ddot{x}} {x}(2\dot{x}^2-x\ddot{x})}+\tan^{-1}\sqrt{\frac{\ddot{x}} {x(2\dot{x}^2-x\ddot{x})}}x\label{int3}.
\end{align}
We can write the general solution of (\ref{eq81}) with the help of the above three integrals and it takes the form
\begin{equation}
x(t)=\sqrt{\frac{I_2} {I_1}}\tan\bigg[\frac{1} {2}(\sqrt{I_1I_2}t+2I_3)\bigg].\label{soln}
\end{equation}
\begin{figure}[ht!]
\centering
\includegraphics[width=0.4\textwidth]{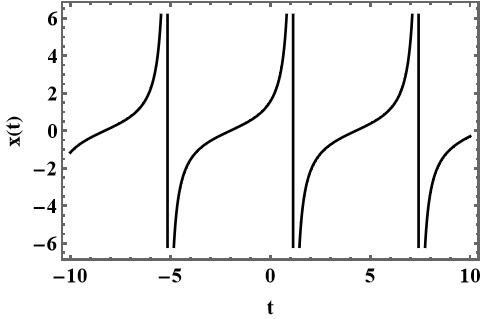}
\caption{Solution plot for Eq. (\ref{eq81}) with the initial conditions $I_1=1,\;I_2=1,\;I_3=1$.}
\label{fig:msf_sig1_sig2_yy_0.3hr}
\end{figure}
It is clear from the above demonstration that one can establish the integrability of the given equation in an algebraic manner  with the help of the procedure developed here.

\section{Conclusion}
In this paper, we have presented a procedure to obtain the integrable quantiifers that are being used in the extended PS procedure from the  DPs and their cofactors.  In this procedure one has to determine a function by solving a first order differential equation. The rest of the procedure involves only algebraic calculations. Interestingly, any particular solution of the aforementioned first order differential equation is sufficient to proceed further and identify the other quantifiers. The main advantage of this interconnection is that the underlying quantifiers can be derived without solving the determining equation in the respective method. Such determination helps in establishing the complete integrability as well as the general solution for the given nonlinear ODE. We have determined the usefulness of the interconnections by considering three different examples.


\section*{Acknowledgments}
The work of M.S. forms part of a research project sponsored by National Board of Higher Mathematics (NBHM), Government of India, under Grant No. 02011/20/2018 NBHM 
(R.P.)/R\&D II/15064.


\end{document}